\documentclass[10pt,a4paper,final]{iopart}
\usepackage{iopams}  
\usepackage{graphicx}
\usepackage[breaklinks=true,colorlinks=true,linkcolor=blue,urlcolor=blue,citecolor=blue]{hyperref}
\usepackage{siunitx}
\usepackage[]{cite}
\usepackage{float}
\usepackage{subfig}
\usepackage{soul}
\begin{document}

\title{Small-signal parameters extraction and noise analysis of CNTFETs}\footnote{This is the Accepted Manuscript version of an article accepted for publication in \emph{Semiconductor Science and Technology}.  IOP Publishing Ltd is not responsible for any errors or omissions in this version of the manuscript or any version derived from it.  The Version of Record is available online at \href{https://iopscience.iop.org/article/10.1088/1361-6641/ab760b}{10.1088/1361-6641/ab760b}.}

\author{Javier N. Ramos-Silva$^{1,2}$, An\'ibal Pacheco-S\'anchez$^{2}$, Mauro A. Enciso-Aguilar$^1$, David Jim\'enez$^2$, Eloy Ram\'irez-Garc\'ia$^{1,3}$}
\address{$^1$ Instituto Polit\'ecnico Nacional, UPALM, Edif. Z-4 3er Piso, Cd. de M\'exico, 07738, M\'exico}
\address{$^2$Departament d'Enginyeria Electr\`onica, Escola d'Enginyeria, Universitat Aut\`onoma de Barcelona, Bellaterra, 08193, Spain}
\address{$^3$ Chair for Electron Devices and Integrated Circuits, Technische Universit\"{a}t Dresden, Dresden, 01069, Germany}
\ead{jramoss1303@alumno.ipn.mx, AnibalUriel.Pacheco@uab.cat, \mailto{ramirezg@ipn.mx}}

\begin{abstract}The use of carbon nanotube (CNT) field-effect transistors (FETs) in microwave circuit design requires an appropriate, immediate and efficient description of their performance. This work describes a technique to extract the parameters of an electrical equivalent circuit for CNTFETs. The equivalent circuit is used to model the dynamic and noise performance at low- and high-frequency of different CNTFET technologies, considering extrinsic and intrinsic device parameters as well as the contact resistance. The estimation of the contact resistance at the metal/CNTs interfaces is obtained from a Y-function based extraction method. The noise model includes four noise sources: thermal noise, thermal channel noise, shot channel noise and flicker noise. The proposed model is compared with a compact model calibrated to hysteresis-free experimental data from a high-frequency multi-tube (MT)-CNTFET technology. Additionally, it has been applied to experimental data from another fabricated MT-CNTFET technology. The comparison in both cases shows a good agreement between reference data (simulation and experimental) and results from the proposed model. Low- and high-frequency noise projections of the fabricated reference device are further studied. Noise results from both studied technologies show that shot noise mainly contributes to the total noise due to the presence of Schottky barriers at contacts and along the channel. 
\end{abstract}

\vspace{2pc}


\section{Introduction}

Nowadays, silicon (\textrm{Si}) based technologies dominate the high-frequency electronic market. However, these components are reaching their performance limits, causing the search for new suitable solutions for high-performance applications, such as transistor technologies based on nanostructured materials. During the last years, carbon nanotube (CNT) field-effect transistors (FETs) have been studied in order to exploit their extraordinary inherent features, e.g., high-linearity \cite{MarRut19,BauPes07,MotCla15,Maa17}, low high-frequency noise \cite{LanGon14,SakSch11}, quasi-ballistic transport and an outstanding gate control. Moreover, some projections refers to operation frequencies in the order of \SI{}{\tera\hertz} \cite{HasSal06,GuoHas05}, as the technological process improve and the limitations due to the fabrication are minimized (e.g., parasitics, growth methods and alignment of CNTs). 

Fabricated CNTFET technologies with an extrinsic operation frequency of around \SI{100}{\giga\hertz} have been reported in the literature \cite{CaoBra16,RutKan19,ZhoShi19}. These improvements have caused research efforts to develop radio-frequency (RF) applications using CNTFETs \cite{RamPac19, SchCla13, EroLin11, TagCar15}. The design of RF circuits based on CNTFET technologies requires a reliable compact model capable to reproduce the novel and unique behavior of the devices when these are used at circuit level computer-aided design. This compact model may help to understand and quantify the effect of parasitic elements on dynamic and noise performance. Also, this knowledge may highlight limitations in the fabrication process of the device, project ideal device performance at different bias conditions and can be useful to benchmark the device versus other technologies in specific scenarios.

The main problem related to the small-signal representation of an emergent transistor technology, such as CNTFETs, is the correct extraction of the parameters for the equivalent circuit, including the non-negligible contact resistances at metal/channel interfaces. As the development of a complete physics-based model for analog applications is challenging due to the presence of physical phenomena, such as potential barriers \cite{SchCla13,SveCam11} produced on the metal/CNT interface, as well as the non-linear behavior of the charge on the tubes \cite{MotCla15,SveCam11}, CNTFET compact modeling  is usually done by a semi-physical \cite{SchHaf15,SchHaf15_2} or by a semi-empirical approach \cite{MarGel15,TuoWan17,KocDun09}, the latter is the approach followed here. 

In this work, the equivalent circuit and the parameters extraction procedure, including the contact resistance extraction, are described in Section \ref{Sec_Eq_Cir}. The low- and high-frequency noise model as well as the description of the noise sources are presented in Section \ref{Sec_Noi_Mod}. The validation of the proposed model using a reference compact model is presented in Section \ref{Sec_Ref_Data}. In Section \ref{Sec_Cha_Dev} the proposed approach is applied to experimental data from a fabricated top-gated MT-CNTFET technology. Finally some conclusions are provided in Section \ref{Sec_Con}. 

\section{Small-signal equivalent circuit and parameters extraction}
\label{Sec_Eq_Cir}
\subsection{Electrical equivalent circuit description}

\begin{figure}[!htb]
	\centering
	{\includegraphics[height=0.35\textwidth]{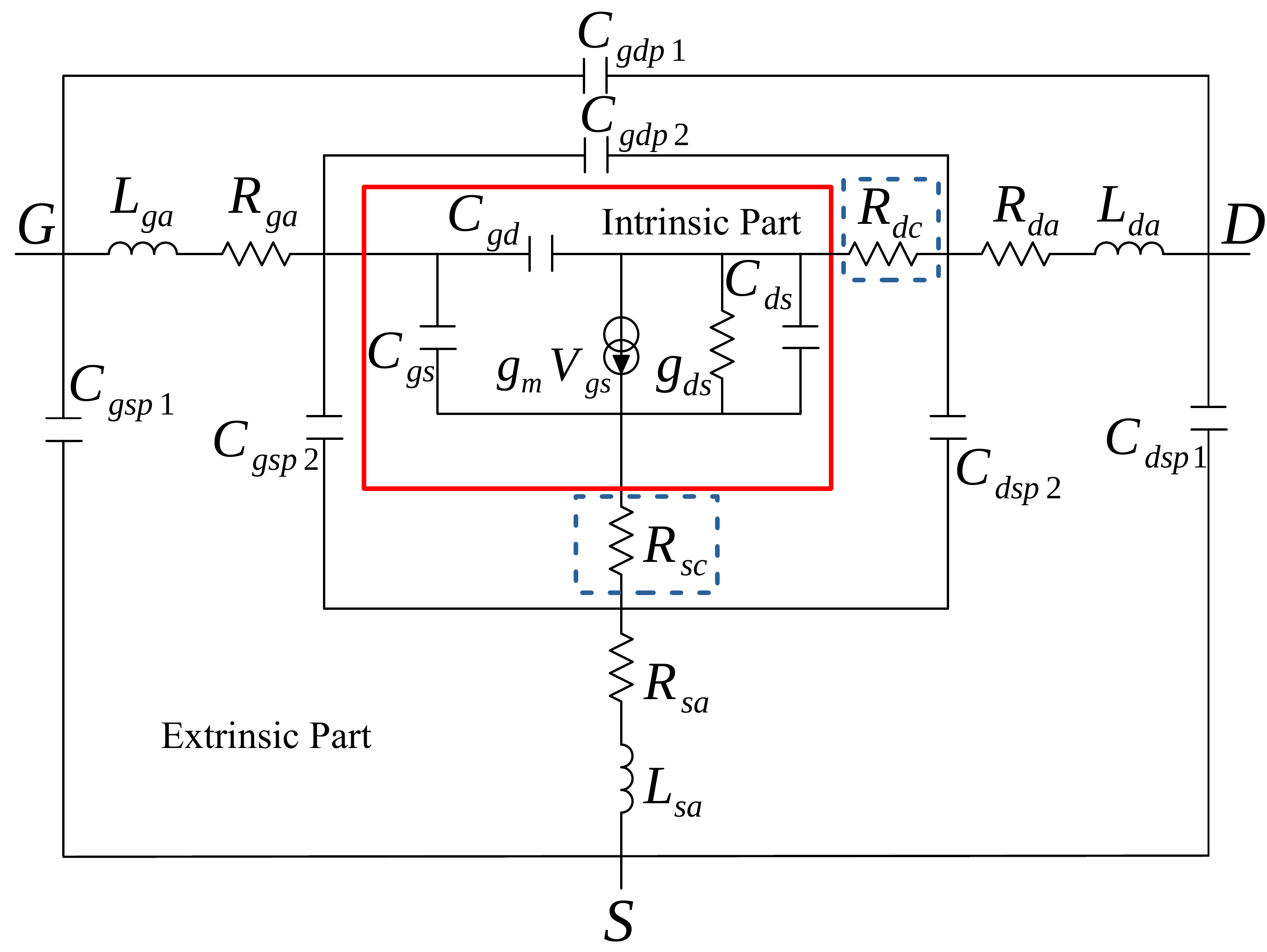}}
	\caption{Equivalent circuit of a CNTFET considering  the contact resistances associated to the metal/CNT interfaces (dashed blue boxes) and the intrinsic parameters (solid red box).}
	\label{fig:02}
\end{figure}

Fig. \ref{fig:02} introduces the equivalent circuit considered in this work, which is divided into three sections. The first one is the extrinsic part, which contains the parasitic parameters of the CNTFET, caused mainly by fabrication process issues, the second one is the intrinsic part, which represents the physical transport phenomena that occur inside the device through electrical parameters and the last one includes the contact resistances, related to the interfaces between CNTs in the channel and the metallic contacts, which are specially large in CNTFETs and can not be neglected due to their large impact in device performance \cite{FraFar14,CaoHan12}. $R_{\rm ga}$, $R_{\rm da}$, $R_{\rm sa}$, $L_{\rm ga}$, $L_{\rm da}$ and $L_{\rm sa}$ are the extrinsic resistances and inductances associated to the access metallizations of the device for the gate, drain and source, respectively. $C_{\rm gdp1}$ and $C_{\rm gdp2}$ are the parasitic capacitances between gate and drain $C_{\rm dsp1}$ and $C_{\rm dsp2}$ between drain and source and $C_{\rm gsp1}$ and $C_{\rm gsp2}$ between gate and source. $R_{\rm dc}$ and $R_{\rm sc}$ are the contact resistances at the drain and source, respectively. In contrast to the parasitic parameters, contact resistances can not be de-embedded using standard procedures and are one of the most important CNTFETs fabrication issues, associated to Schottky barriers at the contacts \cite{SveCam11}, hence, in order to model accurately, the effect of contact resistances must be taken into account. The parameters related to the intrinsic part of the device are described as follows: $g_{\rm ds}$ is the channel or output conductance and $g_{\rm m}$ is the transconductance, $C_{\rm gd}$, $C_{\rm gs}$ and $C_{\rm ds}$ represent the intrinsic capacitances between the gate, drain and source and $V_{\rm gs}$ is the intrinsic voltage between gate and source. 

\begin{figure}[!htb]
	\centering
	{\includegraphics[height=0.31\textwidth]{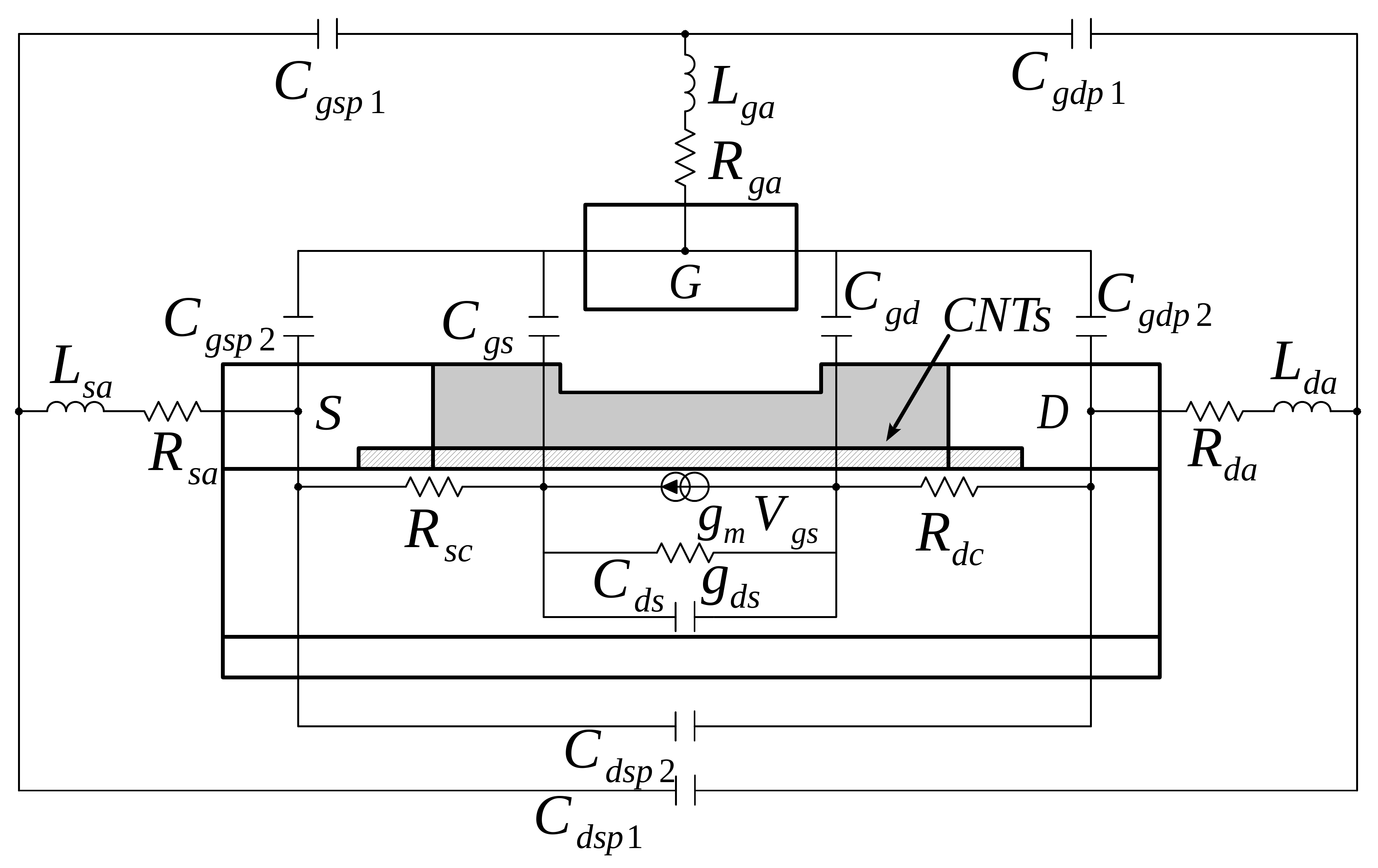}}
	\caption{Parameters of the equivalent circuit associated to the components of the cross-section schematic view for a top-gated CNTFET. The gray area represents high-$k$ oxide and the striped area the CNT-based channel.}
	\label{fig:02_2}
\end{figure}

In Fig. \ref{fig:02_2}, each parameter of the equivalent circuit is associated to the components of the cross-section schematic view for a top-gated CNTFET.

\subsection{Extraction of the extrinsic parameters}

A small output signal of a CNTFET used as an amplifier, an immediate analog application, can be easily affected by the parasitic elements. Thus, it is important to know appropriately the contribution of the extrinsic parameters and separate their contribution from that produced by the intrinsic parameters. 

In order to extract correctly the extrinsic parameters of the device, it is necessary to use an accurate de-embedding procedure. The three-step parasitic de-embedding method is synthesized as in \cite{TieHav03,TieHav05} by equations (\ref{Eq_1}) to (\ref{Eq_4}).

\begin{equation}
\label{Eq_1}
Y_{\rm Dem}=\left(\left(Y_{\rm Raw}-Y_{\rm E}\right)^{-1}-Z_{\rm S}\right)^{-1}-Y_{\rm I},
\end{equation}

\begin{equation}
\label{Eq_2}
Y_{\rm Open}=\left(\left(Y_{\rm I}\right)^{-1}+Z_{\rm S}\right)^{-1}+Y_{\rm E},
\end{equation}

\begin{equation}
\label{Eq_3}
Y_{\rm Short}=\left(Z_{\rm S}^{-1}\right)+Y_{\rm E},
\end{equation}

\begin{equation}
\label{Eq_4}
Y_{\rm Pad}=Y_{\rm E},
\end{equation}

\noindent where $Y_{\rm Open}$, $Y_{\rm Short}$ and $Y_{\rm Pad}$ are the $Y$-parameters of an open, short and thru test structure, respectively. $Y_{\rm E}$, $Y_{\rm I}$ and $Z_{\rm S}$ are the admittance outer-parasitics matrix, the admittance inner-parasitics matrix and the series-impedance parasitics matrix, respectively, (see Fig. 2 and eqs. (17) to (19) in \cite{TieHav05}). $Y_{\rm Dem}$ represents the de-embedded $Y$-parameters. The three-step parasitic de-embedding method has been considered suitable to be applied to CNTFET characterization due to the high operation frequency expected in this technology \cite{HasSal06,GuoHas05,RutKan19,ZhoShi19,CaoBra16} where an appropriate de-embedding method can improve accuracy of the device experimental data \cite{TieHav05,KooGee91}.

Notice that the extracted extrinsic parameters do not consider the contribution of the contact resistances at the source and drain sides since these resistances are produced on the interface between CNTs and contact metallizations. Therefore, in order to have an accurate model for CNTFETs, the contribution of contact resistances is required to be removed by an additional process.

\subsection{Contact resistance extraction}

The total contact resistance ($R_{\rm C}=R_{\rm sc}+R_{\rm dc}$) is an electrical representation of the physical phenomena at the junction metal/CNT at both drain and source contacts. This is one of the main issues that limits CNTFETs performance, thus it has to be considered in an accurate model. Also, an appropriate extraction can provide information about the contact quality to improve fabrication process and to project the performance of an optimized technology.

In order to perform an accurate extraction of the contact resistance of CNTFETs, a Y-function based method, labeled as YFM$_{\rm 2}$ in \cite{PacCla16}, has been considered\footnote{Notice that the Y-function is defined as $Y=I_Dg_m^{(-1/2)}$ and has no relation with the admittance parameters discussed along this manuscript.}. In contrast to the transfer length method (TLM), YFM$_{\rm 2}$ does not require long CNT test structure \cite{FraFar14} and only standard transfer characteristics in the linear region and at low drain bias of the device are required to extract $R_{\rm C}$. Also, a physics-based validation has been provided for this method in \cite{PacCla16}. Despite the extracted resistance is not bias dependent, it can be considered as a reference value for near bias points as well as an upper limit value of $R_{\rm C}$ for the device working in the active region. 

As is stated in \cite{PacCla16}, in YFM$_{\rm 2}$ the $R_{\rm C}$ is extracted from the slope of $\theta=\theta_0+R_{\rm C}\beta$ once $\beta$, $\theta$ and $\theta_0$, the degradation factor of low-field carrier mobility and the extrinsic and intrinsic reduction mobility carrier coefficient, respectively, are extracted from the Y-function obtained from the drain current equation described in \cite{PacCla16} (see eqs. 3 and 9 in \cite{PacCla16}).

\subsection{Intrinsic parameters extraction}
\label{sec_extra_intri}

By considering the $\Pi$-like topology of the intrinsic part of the equivalent circuit shown in Fig. \ref{fig:02}, the admittance $Y$-parameters have been used to characterize the two port network representation of the CNTFET as follows:

\begin{eqnarray}
\nonumber Y_{\rm i}(\omega)&=\left[
\begin{array}{cc}
y_{\rm 11,i} & y_{\rm 12,i} \\
y_{\rm 21,i} & y_{\rm 22,i} \\ 
\end{array}
\right]
\\
&=\left \{ \begin{array}{cc}
y_{\rm 11,i}=i\omega(C_{\rm gs}+C_{\rm gd}),\\
y_{\rm 12,i}=-i\omega C_{\rm gd},\\
y_{\rm 21,i}=g_{\rm m}-i\omega C_{\rm gd},\\
y_{\rm 22,i}=g_{ds}+i\omega(C_{\rm gd}+C_{\rm ds}),
\end{array}
\right.
\end{eqnarray}

\noindent from which the intrinsic parameters can be obtained: 

\begin{equation}
\label{Eq_gm}
g_{\rm m}=\textrm{Re}\{y_{\rm 21,i}\},
\end{equation}

\begin{equation}
\label{Eq_gds}
g_{\rm ds}=\textrm{Re}\{y_{\rm 22,i}\},
\end{equation}

\begin{equation}
\label{Eq_Cds}
C_{\rm ds}=\frac{\textrm{Im}\{y_{\rm 22,i}\}}{\omega}-C_{\rm gd},
\end{equation}

\begin{equation}
\label{Eq_Cgd}
C_{\rm gd}=C_{\rm gs}=-\frac{\textrm{Im}\{y_{\rm 12,i}\}}{\omega}.
\end{equation}

Notice that the device electrostatics is considered symmetrical, i.e., $C_{\rm gd}$ is identical to $C_{\rm gs}$.

\begin{figure}[!htb]
	\centering
	{\includegraphics[height=0.24\textwidth]{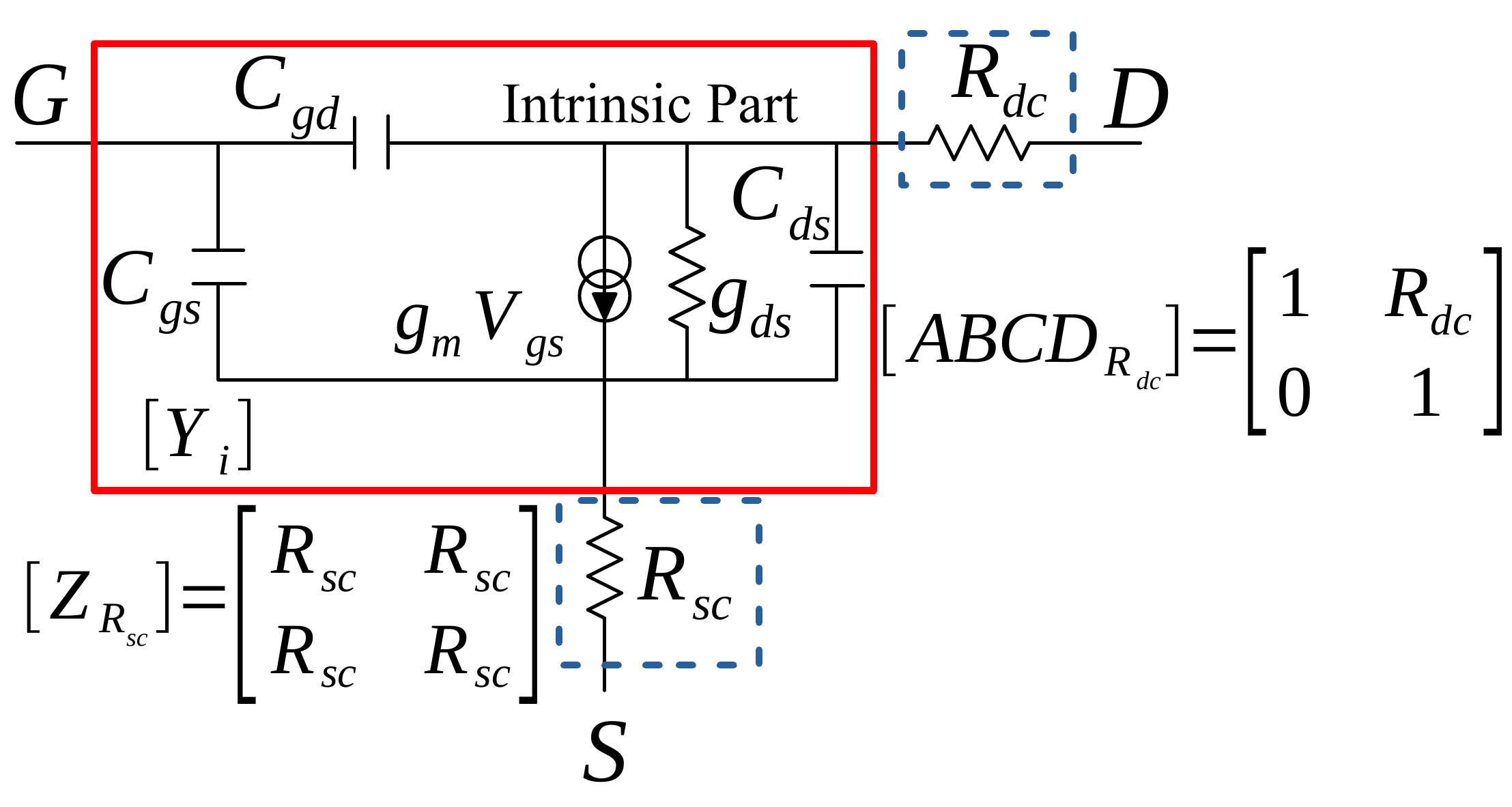}}
	\caption{Intrinsic part of the equivalent circuit  and contact resistances matrix expressions used to extract the intrinsic $Y$-parameters from de-embedded $Y$-parameters.}
	\label{fig:second}
\end{figure}

In contrast to previous works \cite{KocDun09,MarGel15,TuoWan17}, the approach used here is based on linear matrices in order to remove the contribution of the contact resistances from the de-embedded parameters. As Fig. \ref{fig:second} shows, $\left[Z_{\rm R_{\rm sc}}\right]$ is connected in series with the intrinsic part $\left[Y_{\rm i}\right]$, while $\left[ABCD_{\rm R_{\rm dc}}\right]$ is connected in a cascade topology. 

Then, using the following equations, where the operator $\rightarrow$ implies a matrix transformation, intrinsic admittance parameters can be obtained from de-embedded parameters.  

\begin{equation}
\left[Z_{\rm i,R_{\rm dc}}\right]=\left[Z_{\rm Dem}\right]-\left[Z_{\rm R_{\rm sc}}\right],
\end{equation}

\begin{equation}
\left[Z_{\rm i,R_{\rm dc}}\right]\rightarrow\left[ABCD_{\rm i,R_{\rm dc}}\right],
\end{equation}

\begin{equation}
\left[ABCD_{\rm i}\right]=\frac{\left[ABCD_{\rm i,R_{\rm dc}}\right]}{\left[ABCD_{\rm R_{\rm dc}}\right]},
\end{equation}

\begin{equation}
\left[ABCD_{\rm i}\right]\rightarrow\left[Y_{\rm i}\right].
\end{equation}

Once $\left[Y_{\rm i}\right]$ is known, intrinsic parameters can be obtained directly from equations (\ref{Eq_gm}) to (\ref{Eq_Cgd}). 

\section{Low- and high-frequency noise model for CNTFETs}
\label{Sec_Noi_Mod}

Four different noise sources have been considered in this work: thermal noise associated to resistive elements, shot channel noise due to the Schottky junctions, thermal channel noise associated to an equivalent noise temperature and flicker noise (1/$f$), mostly present at low frequencies. Details related to each noise source are presented below. 

\subsection{Thermal noise in resistive elements}

The thermal noise in resistive elements of the equivalent circuit, as a result of the movement of the free carriers inside the conductive material \cite{Ziel86}, can be described by its power spectral density as:

\begin{equation}
\label{ruidot}
S_{\rm T}=\frac{\bar{i}^2_{\rm T}}{\Delta f}=\frac{4 k_{\rm B} T}{R},
\end{equation}

\noindent where $k_{\rm B}$ is the Boltzmann constant, $R$ is the resistance, $\Delta f$ is the noise bandwidth, which is considered equal to one for all cases in this work and $T$ is the absolute temperature. Each resistance present in the equivalent circuit has a characteristic power spectral density to describe its produced thermal noise. 

\subsection{Shot channel noise}

The carrier injection in CNTFETs will be affected as a consequence of the potential barriers in the channel \cite{NavJun07,LanGon12,LanGon14}. The carrier injection through the potential barrier, when a low carrier density is present in the channel, follows a Poisson's probability distribution. The result is a power spectral density associated to the shot noise represented in equation (\ref{shot}) where $q$ is the electron charge, $I_{\rm D}$ is the drain current and $F$ the Fano Factor, which represents the compression of the shot noise as a result of the correlation between successive carriers injections in the potential barriers \cite{BetFio09}. When no correlation exists $F$ is equal to one, as it has been considered in this work.          

\begin{equation}
\label{shot}
S_{\rm Shot}=\frac{\bar{i}^2_{\rm Shot}}{\Delta f}=2qI_{\rm D}F.
\end{equation}

\subsection{Thermal channel noise}

The relaxed channel length (e.g., $L_{\rm ch}>\SI{300}{\nano\meter}$) \cite{SchKol11,CaoBra16,MarRut19} and the large density of tubes in high-frequency CNTFETs, as well as the bias near saturation required for the dynamic performance of these devices, induce the activation of scattering mechanisms of acoustic and optical phonons, hence, non-ideal ballistic transport exists in these devices and the channel resistance can not be neglected \cite{PacCla16}. As a result of this channel resistance, thermal channel noise spectral power density is considered as in the Pospieszalski model \cite{Pos89}. The corresponding expression, based on drift-diffusion theory and applied here to CNTFETs due to the transport conditions described above, is shown in equation (\ref{tch}).   

\begin{equation}
\label{tch}
S_{\rm TCh}=\frac{\bar{i}^2_{\rm TCh}}{\Delta f}=4k_{\rm B}T_{\rm d}g_{\rm ds}.
\end{equation}

An equivalent noise temperature ($T_{\rm d}$), associated to the output conductance, is needed to describe the thermal channel noise. Equation (\ref{td}) describes the equivalent noise temperature $T_{\rm d}$\cite{AguCro05}, as it can be noticed, $T_{\rm d}$ is independent to the frequency and only depends on the room temperature and the relation between $g_{\rm m}$ and $g_{\rm ds}$. 

\begin{equation}
\label{td}
T_{\rm d}=\left(1+\frac{g_{\rm m}}{g_{\rm ds}}\right)T_{\rm amb}.
\end{equation}

Considering a model for thermal noise in the channel explicitly for CNTFETs in this study contrasts with other approaches where such phenomenon is not considered \cite{LanGon14,SakSch11} or is implicit in the description of the transport through a transmission probability \cite{SchHaf15,SchHaf15_2}.

\subsection{Flicker noise (1/$f$)}

Flicker noise presents an 
inversely proportional power spectral density as a function of the frequency. Previous studies refer that flicker noise is specially large in CNTFETs \cite{LanGon12,LinApp06,ColFuh00,AppLin07} and it is related to the total number of charge carriers \cite{LinApp06}.

The power spectral density of the flicker noise is described in equation (\ref{flicker}). In this equation $A_{\rm H}$ represents the noise amplitude which scales the noise contribution and is equal to the relation between the Hooge's constant ($a_{\rm H}$), dependent on the scattering channel mechanisms, and the number of charge carriers in the channel ($n$) \cite{Hoo69}. On the other hand, $a_{\rm H}$ is an empirical parameter which is commonly in the order between  $10^{-4}$ and $10^{-3}$ for non-optimized materials \cite{Hoo94} as is the CNTFETs case. For this work the chosen value was $a_{\rm H}=2\times 10^{-4}$. Notice that a dominant mobility variation mechanism has been considered here to describe the flicker noise.

\begin{equation}
\label{flicker}
S_{1/f}=\frac{\bar{i}^2_{1/f}}{\Delta f}=A_{\rm H}\frac{I_{\rm D}^2}{f}=\frac{a_{\rm H}}{n}\frac{I_{\rm D}^2}{f}.
\end{equation}

\begin{figure}[!htb]
	\centering
	{\includegraphics[height=0.27\textwidth]{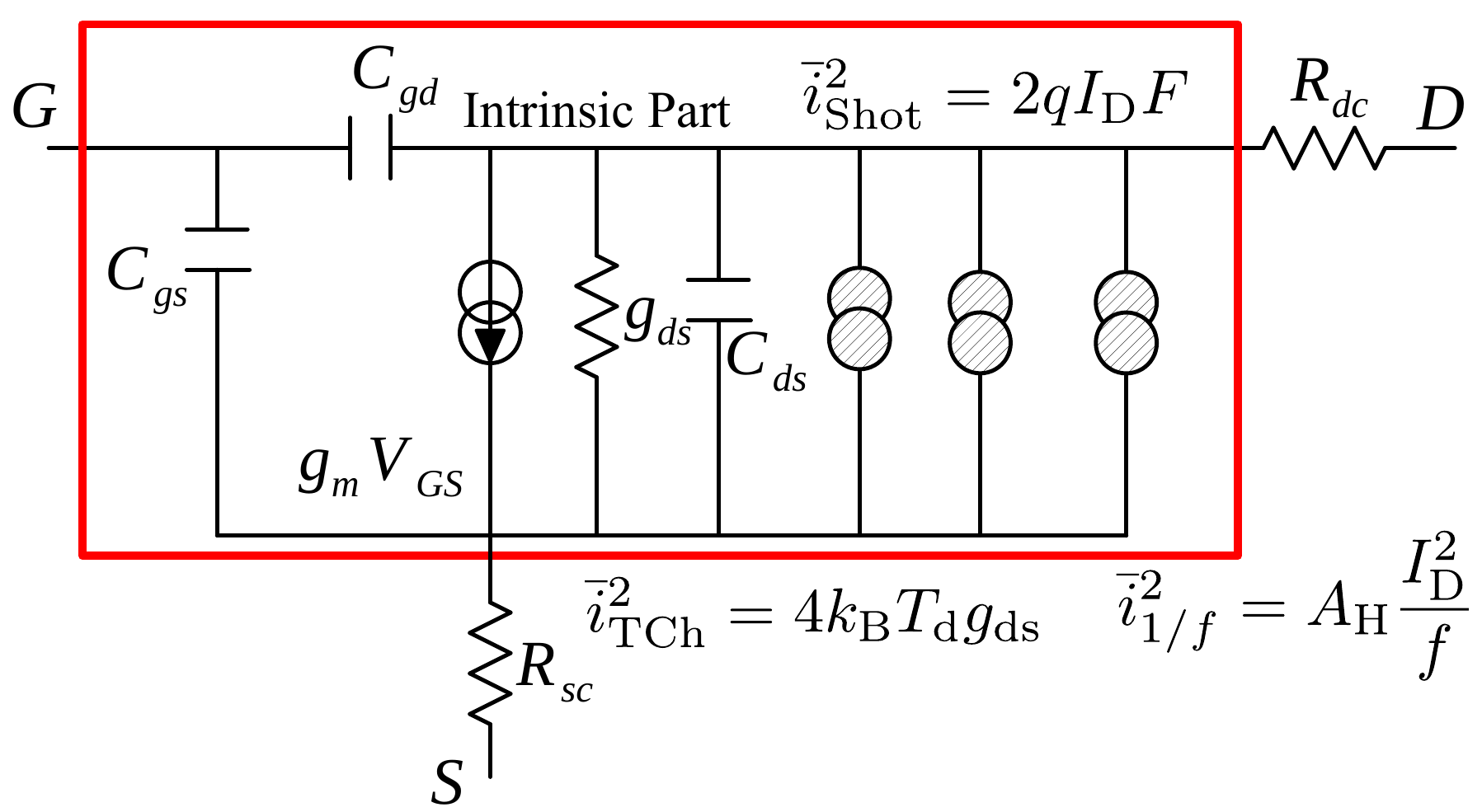}}
	\caption{Intrinsic part of the equivalent circuit with the noise sources: shot noise, thermal channel noise and $1/f$ noise. Thermal noise is considered in all resistive parameters of the equivalent circuit shown in Fig. \ref{fig:02_2}.}
	\label{ruido}
\end{figure}

Once that all proposed noise sources are known these can be added to the electrical equivalent circuit as shows Fig. \ref{ruido}.

As it can be noticed, the considered noise model is related to the small-signal model presented before and as a result of that, accurate projections for dynamic and noise CNTFETs performance are only possible if all parameters are extracted appropriately. Thus, a correct extraction of contact resistance is mandatory in order to achieve a good approximation to CNTFETs performance.

\section{Validation of the proposed model using a reference compact model}
\label{Sec_Ref_Data}
\subsection{Initial reference data}

The compact model used as a reference in this work \cite{SchHaf15,SchHaf15_2}, which is called CCAM, has been originally calibrated to hysteresis-free experimental data from a CNTFET technology presented elsewhere \cite{SchKol11}. Specifically, the top-gated device used as a reference for this work has a device channel length ($L_{\rm ch}$) of around \SI{700}{\nano\meter} and a gate length  ($L_{\rm g}$) of around \SI{250}{\nano\meter}. The device has eight gate fingers of \SI{50}{\micro\meter} each one forming a total device width ($w_{\rm g}$) of approximately \SI{400}{\micro\meter}. The semiconducting-to-metallic (s:m) CNT ratio is 3:1 with around 3000 CNTs randomly distributed in channel and grown via chemical vapor deposition (CVD) on SiO$_{2}$. Metallic contacts of the device are made from Pd/Ti. Details on the fabrication process can be found in \cite{SchKol11}. 

\begin{figure}[!htb]
	\centering
	{\includegraphics[height=0.25\textwidth]{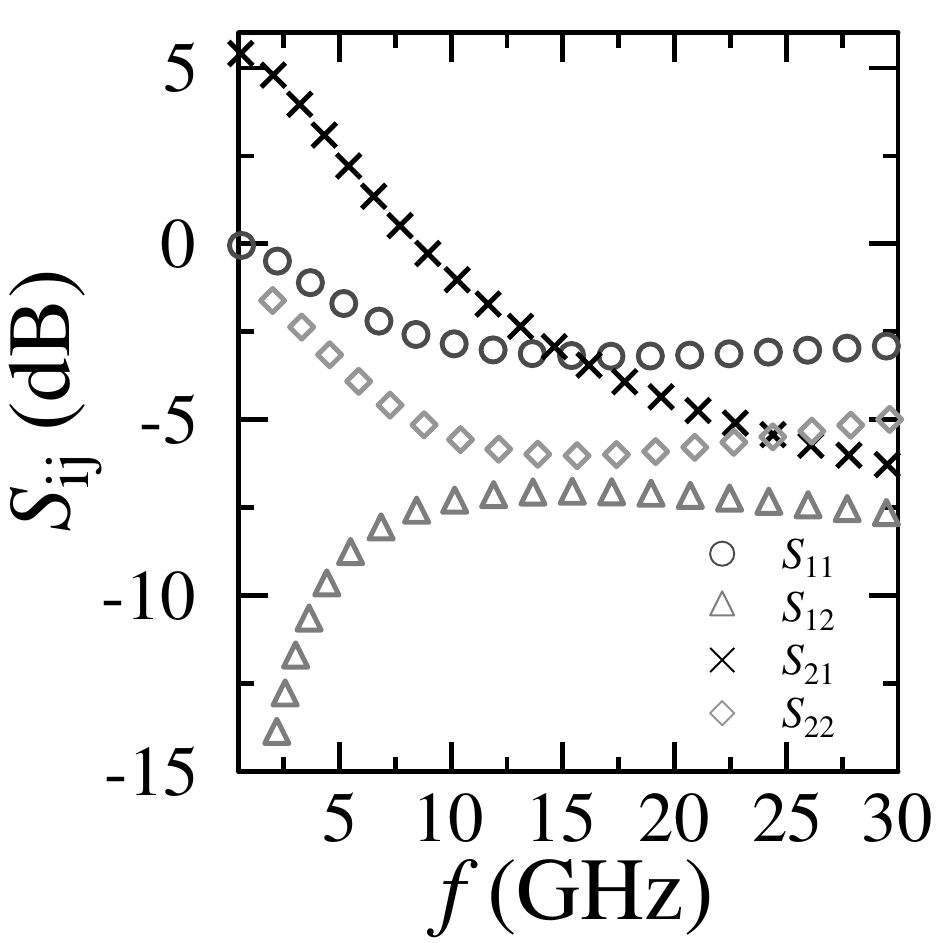}}
	\caption{$S$-parameters of the reference technology obtained from CCAM at $V_{\rm GS}=\SI{1}{\volt}$ and $V_{\rm DS}=\SI{3}{\volt}$.}
	\label{fig:01}
\end{figure}

In order to perform the extraction of the associated equivalent circuit parameters, results obtained from the reference compact model have been used. The contribution of metallic tubes has been turned-off for this study in order to work with the best-case scenario. The bias point has been chosen at $V_{\rm GS}=\SI{1}{\volt}$ and $V_{\rm DS}=\SI{3}{\volt}$ since an optimal high-frequency performance has been found out for this technology under such conditions \cite{RamPac19}. The transit frequency ($f_{\rm t}$) and maximum oscillation frequency ($f_{\rm max}$) obtained with the reference model for the considered DC bias point are \SI{9.8}{\giga\hertz} and \SI{27.9}{\giga\hertz}, respectively. By using CCAM, the $S$-parameters of the device have been simulated from \SIrange{1}{30}{\giga\hertz} as shown in Fig. \ref{fig:01}. These results have been used as the starting point to extract the parameters of the equivalent circuit (see Fig. \ref{fig:02}) for this device by using the methodology presented in Section \ref{Sec_Eq_Cir}. 

\subsection{Parameters extraction results}

In this case, the extrinsic parameters have been obtained directly from CCAM, where access resistances and inductances have been neglected, since pad-de-embedding data is used for the original calibration \cite{SchHaf15}. The extrinsic parameters are summarized in Table \ref{Tab_1}.

\begin{figure}[!htb]
	\centering
	{\includegraphics[height=0.25\textwidth]{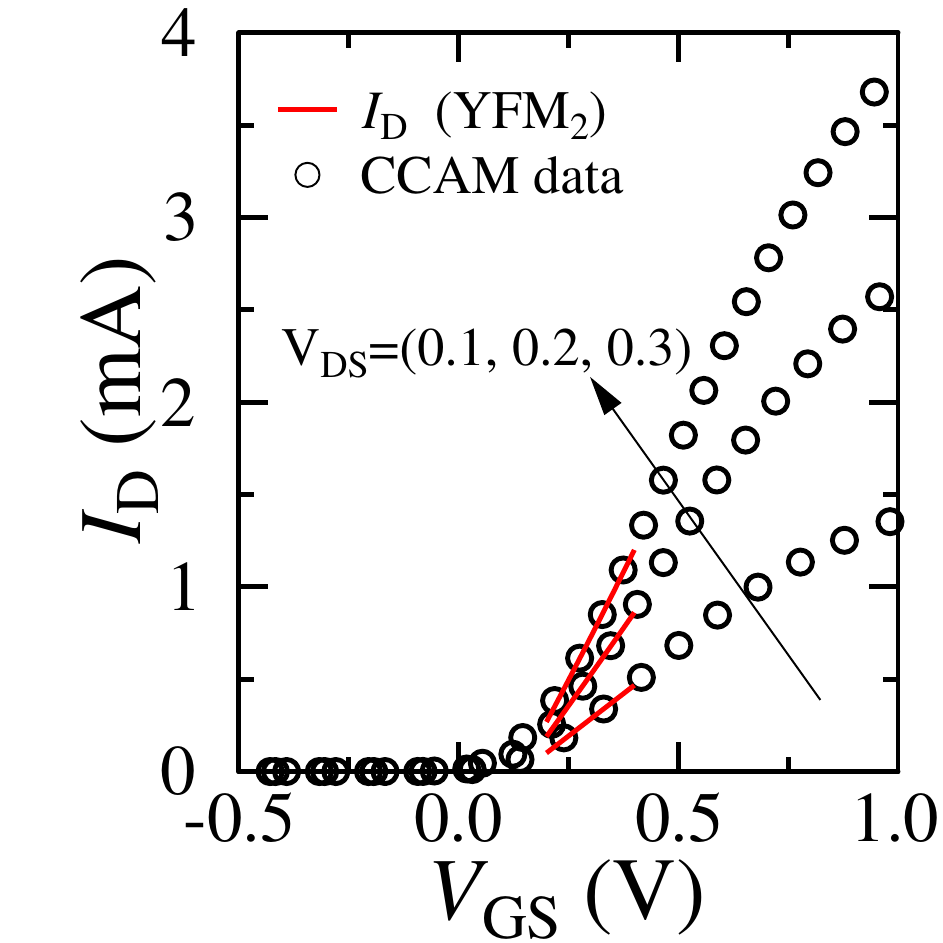}}
	\caption{Transfer characteristics of the device at low drain voltage simulated with CCAM (black markers) and the drain current calculated using the extracted values from YFM$_{2}$ considering $R_{\rm C}\sim\SI{39}{\ohm}$ (solid red lines).}
	\label{fig:03_2}
\end{figure}

Regarding to the contact resistance extraction, Fig. \ref{fig:03_2} shows the transfer characteristics of the device simulated with the reference compact model. Red lines in Fig. \ref{fig:03_2} represent the drain current calculated with a drift-diffusion approach (see Eq. (3) in \cite{PacCla16}) considering $R_{\rm C}\sim\SI{39}{\ohm}$, extracted using YFM$_{\rm 2}$. 

The device exhibit an $n$-type behavior, i.e., carriers are injected from the source to the channel, which causes a larger impact of the potential barrier at the source on the device performance. Considering the latter, a distribution of the total contact resistance is proposed as follows: $R_{\rm dc}=\SI{12}{\ohm}$ and $R_{\rm sc}=\SI{27}{\ohm}$. These values are confirmed by a fitting procedure involving the noise performance developed in Section \ref{Sec_Res_Dis}.

Then intrinsic parameters have been extracted following the proposed procedure (see Section \ref{sec_extra_intri}). Parameters of the extracted equivalent circuit are summarized in Table \ref{Tab_1}.

\begin{table}[!hbt]
	\caption{\label{Tab_1}Comparison of the extracted and optimized values of the equivalent circuit.}
	\begin{indented}
		\item[]\begin{tabular}{@{}lll}
\br
Parameter&Extracted value&Optimized value\\
\mr
\centre{3}{Intrinsic part}\\
\mr
\ns
$g_{\rm m}$	(\SI{}{\milli\siemens})	&   \SI{47.93}{}    &   \SI{47.93}{}    \\ 
$g_{\rm ds}$ (\SI{}{\milli\siemens})&  \SI{2.89}{}     &  \SI{2.89}{}    \\ 
$C_{\rm gd}$ (\SI{}{\femto\farad})	&  \SI{104.12}{}     &  \SI{104.12}{}     \\ 
$C_{\rm gs}$ (\SI{}{\femto\farad})	&  \SI{104.12}{}     & \SI{104.12}{}         \\
$C_{\rm ds}$ (\SI{}{\femto\farad})&  \SI{27.31}{}     &  \SI{27.31}{} \\
\mr
\centre{3}{Contact resistance}\\
\mr
$R_{\rm sc}$ (\SI{}{\ohm})	& \SI{27}{}      &  \SI{27}{}     \\
$R_{\rm dc}$ (\SI{}{\ohm})& \SI{12}{}      &  \SI{12}{}    \\ 
\mr
\centre{3}{Extrinsic part}\\
\mr
$C_{\rm gdp1}$ (\SI{}{\femto\farad})	& \SI{30}{}      &  \SI{30}{}      \\ 
$C_{\rm gdp2}$ (\SI{}{\femto\farad}) & \SI{30}{}      &  \SI{30}{}     \\ 
$C_{\rm dsp1}$ (\SI{}{\femto\farad})	& \SI{20}{}      &  \SI{30}{}     \\ 
$C_{\rm dsp2}$ (\SI{}{\femto\farad})	& \SI{0}{}      &   \SI{0}{}     \\ 
$C_{\rm gsp1}$ (\SI{}{\femto\farad})	& \SI{0}{}      &  \SI{70}{}     \\ 
$C_{\rm gsp2}$ (\SI{}{\femto\farad})	& \SI{80}{}      &  \SI{80}{}     \\
\br
\end{tabular}
\end{indented}
*Extracted extrinsic parameters have been obtained directly from CCAM calibration data. Access resistances and inductances have been neglected during CCAM calibration process. 
\end{table}

In order to demonstrate the feasibility of the extracted transconductance and output conductance, these have been compared with results from equations for traditional FETs considering a non-negligible contact resistance with no bias-dependence (equations (\ref{gm}) and (\ref{gd})). Such equations have described properly different FET technologies, regardless the channel length, in saturation and linear regimes \cite{ChoAnt87}.

\begin{equation}
\label{gm}
g_{\rm m}=\frac{g_{\rm m}'}{1-g_{\rm m}'R_{\rm sc}-g_{\rm ds}'(R_{\rm sc}+R_{\rm dc})}, 
\end{equation}

\begin{equation}
\label{gd}
g_{\rm ds}=\frac{g_{\rm ds}'}{1-g_{\rm m}'R_{\rm sc}-g_{\rm ds}'(R_{\rm sc}+R_{\rm dc})}.
\end{equation}

$g_{\rm m}'$ and $g_{\rm ds}'$ can be obtained from de-embedded data ($Y_{\rm Dem}$, see eq. (\ref{Eq_1})) using equations (\ref{Eq_gm}) and (\ref{Eq_gds}), respectively. In this case, as a result of the hysteresis-free data, $g_{\rm m}'$ and $g_{\rm ds}'$ can also be obtained from DC curves. A comparison between extracted and calculated results of $g_{\rm m}$ and $g_{\rm ds}$ shows that both results are equal, therefore both approaches to obtain these intrinsic parameters are valid. 

Notice that the charge and current transport are described in this work by relations based on the admittance device parameters (see Eqs. (\ref{Eq_gm}) to (\ref{Eq_Cgd})) in contrast to CCAM where such phenomena have been described by semi-empirical expressions. Hence, the approach presented here is expected to be more efficient for immediate characterization purposes.

\subsection{Results and discussion}
\label{Sec_Res_Dis}

The corresponding simulation of the electrical equivalent circuit has been implemented using Keysight Advanced Design System (ADS) from the extracted values of the Table~\ref{Tab_1}. The considered bias point is $V_{\rm GS}=1\ \rm V$ and $V_{\rm DS}=3\ \rm V$ resulting in $I_{\rm D}=\SI{0.015}{\ampere}$. 

At frequencies up to $f_{\rm t}$, which is around \SI{9.8}{\giga\hertz}, the performance of $S$-parameters is similar for both CCAM and the extracted equivalent circuit. However, at frequencies higher than $f_{\rm t}$, the difference between results from CCAM and from the equivalent circuit increases (not shown here). Due to this difference, mainly caused by extrinsic parameters, an optimization has been performed in order to find a good agreement. The optimization procedure consists in minimizing the difference between dynamic and noise results from the proposed model and reference data by fitting only extrinsic parameters, this has been done using the optimization tools of ADS based on a least-squares error-function \cite{Lau85,RouEsc95}. Extracted and optimized parameters for the electrical equivalent circuit are compared in Table \ref{Tab_1}.

\begin{figure}[!htb]
	\centering
	\subfloat[]{\includegraphics[height=0.24\textwidth]{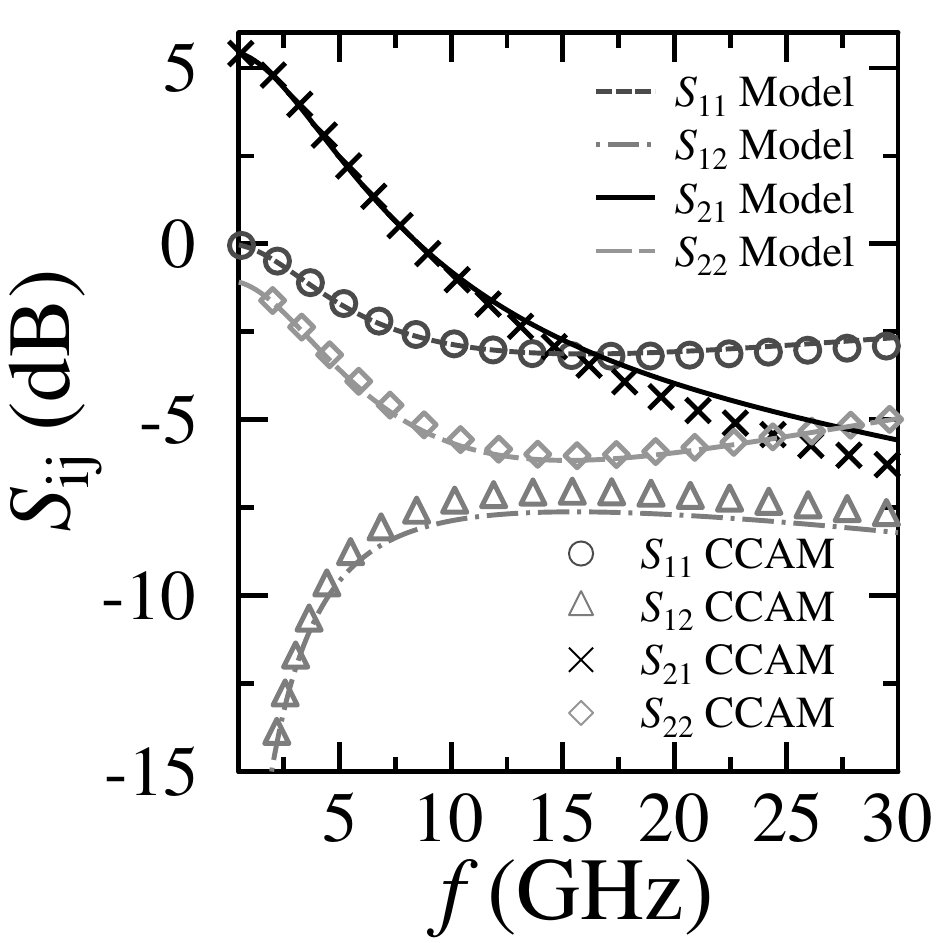}}
	\subfloat[]{\includegraphics[height=0.24\textwidth]{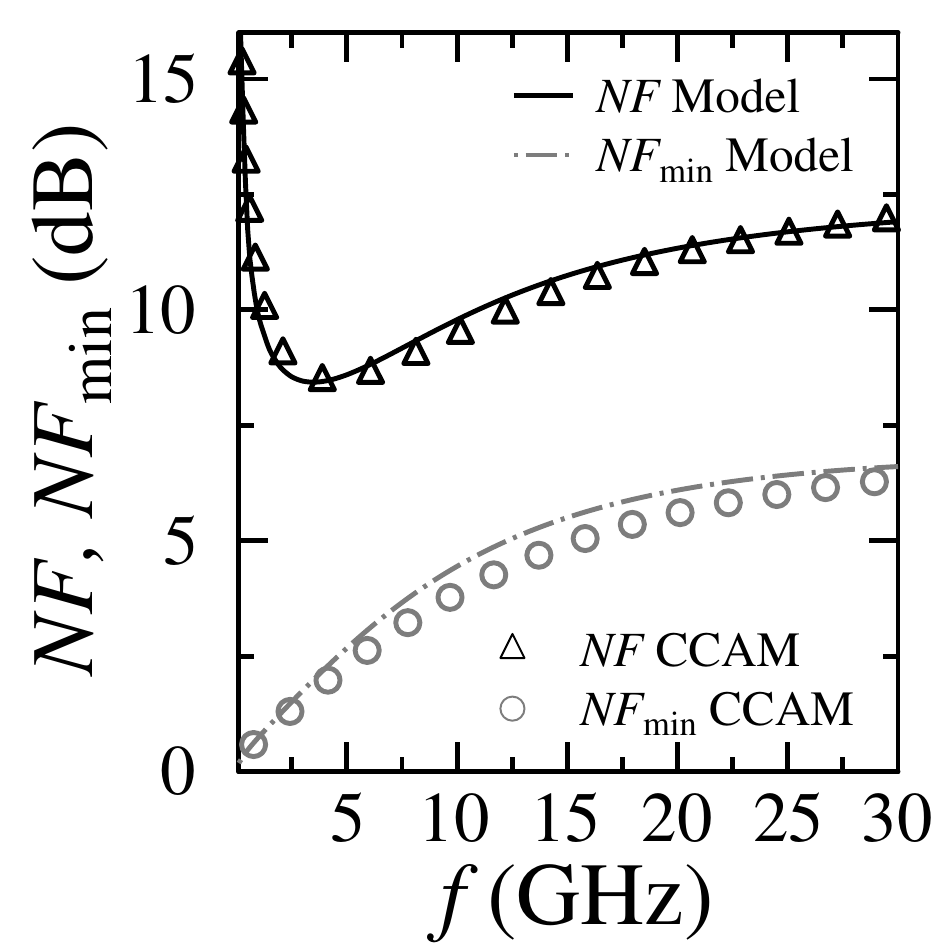}}
	\caption{Comparison of the (a) $S$-parameters and (b) $NF$ and $NF_{\min}$, results from CCAM (markers) and results of the proposed model considering optimized parameters (lines) for $V_{\rm GS}=1\ \rm V$ and $V_{\rm DS}=3\ \rm V$.}
	\label{Ajuste}
\end{figure}

Table \ref{Tab_1} shows that only two extrinsic parameters have changed as a result of the optimization process. A good agreement has been found for results from CCAM and from the proposed equivalent circuit for $S$-parameters as well as for noise parameters (only $NF$ and $NF_{\rm min}$ are shown here) when optimized values are considered as shown in Figs. \ref{Ajuste}(a) and \ref{Ajuste}(b). A correct distribution of the total contact resistance allows a good agreement of $S$-parameters and noise results from equivalent circuit and from CCAM. From Fig. \ref{Ajuste}(b) it can be observed that a good agreement between equivalent circuit and CCAM has been found for frequencies commonly used for circuit design  @$2.4/\SI{5}{\giga\hertz}$: for the proposed model $NF=8.58/\SI{8.59}{\decibel}$ and $NF_{\rm min}=1.46/\SI{2.67}{\decibel}$ while for CCAM $NF=8.84/\SI{8.47}{\decibel}$ and $NF_{\rm min}=1.29/\SI{2.29}{\decibel}$.

\begin{figure}[!htb]
	\centering
	\subfloat[]{\includegraphics[height=0.24\textwidth]{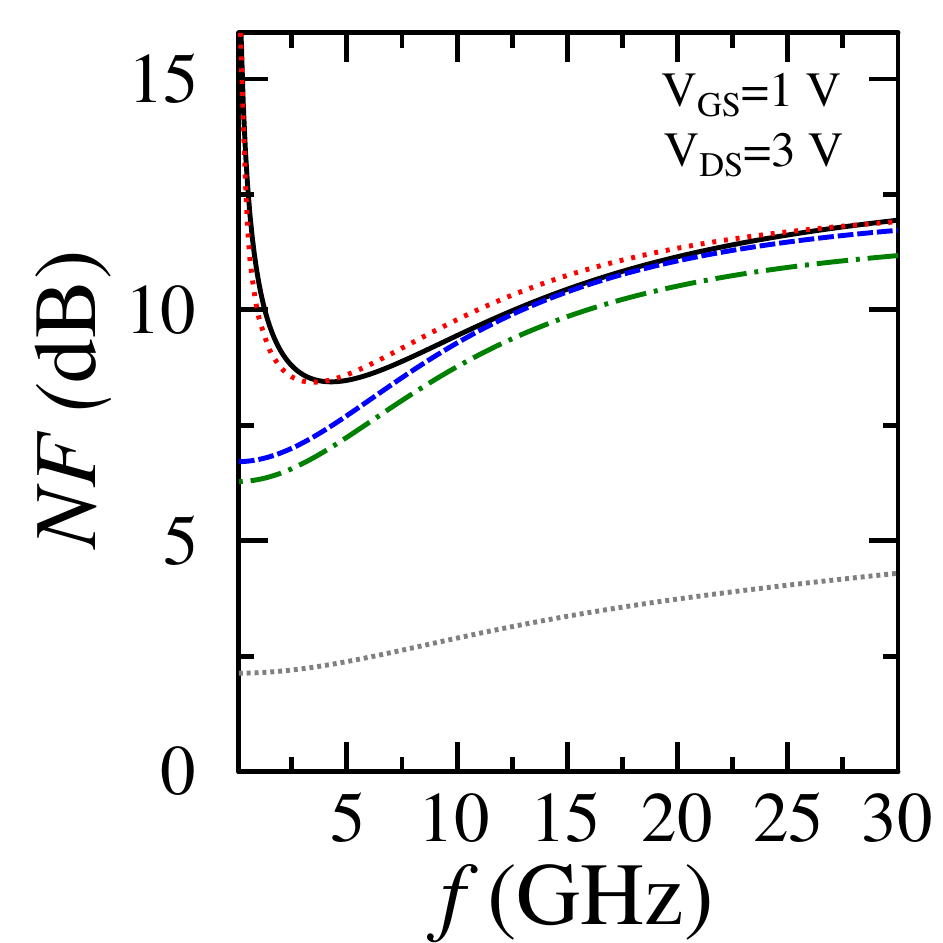}}
	\subfloat[]{\includegraphics[height=0.24\textwidth]{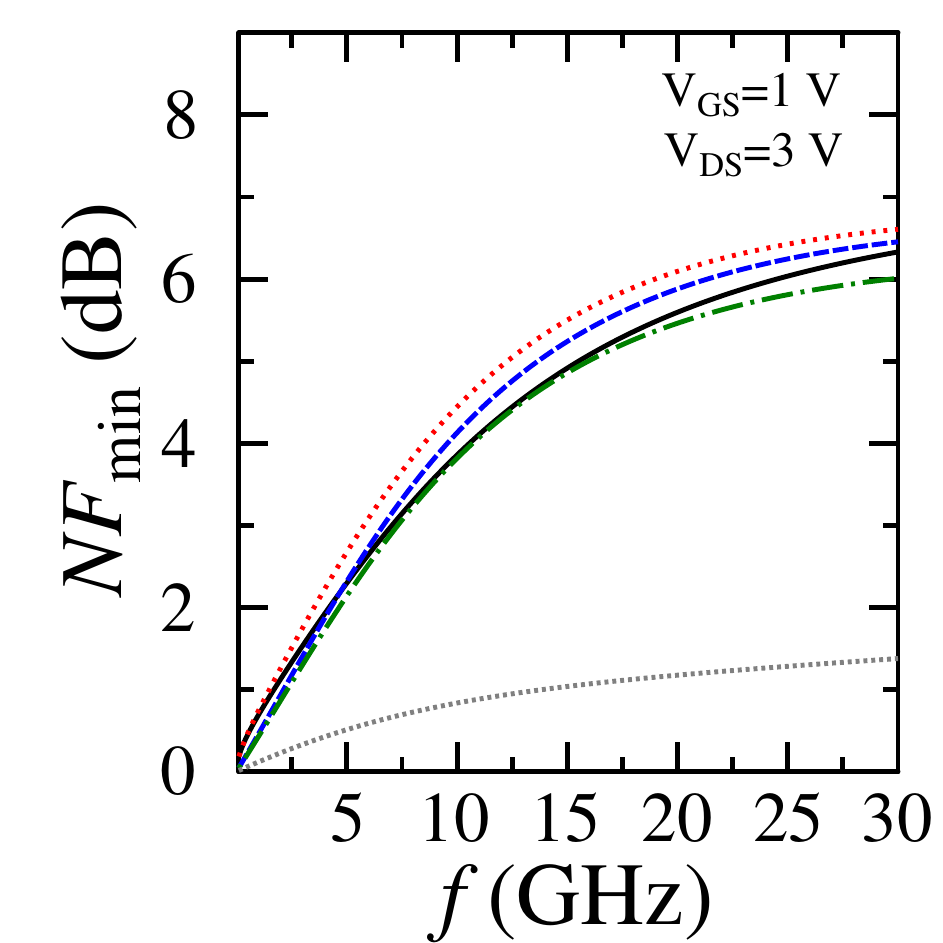}}
	\caption{(a) $NF$ and (b) $NF_{\rm min}$ for: CCAM (solid black lines), equivalent circuit with the four noise sources (dotted red lines in the electronic version), without $1/f$ noise (cut blue lines  in the electronic version), without $1/f$ nor thermal channel noise (cut-dotted green lines in the electronic version) and without $1/f$ noise nor thermal channel noise nor shot noise (dotted gray lines).}
	\label{Noise_Fuentes}
\end{figure}

From noise results depicted in Fig. \ref{Noise_Fuentes} it has been shown that shot noise mainly contributes to the total noise over all frequency range. This trend has been reported in \cite{SakSch11,LanGon12,MarGel17}, while flicker noise does at low frequencies, the same trend has been found in \cite{LanGon12,LinApp06,AppLin07,ColFuh00}.

\section{Characterization of a fabricated device}
\label{Sec_Cha_Dev}

The proposed methodology and model have been also applied to experimental data from a fabricated top-gated MT-CNTFET which is detailed in \cite{KocDun09}. Extrinsic transit frequency ($f_{\rm t,e}$) and extrinsic maximum oscillation frequency ($f_{\rm max,e}$) obtained from measured $S$-parameters are $\sim\SI{5}{\giga\hertz}$ and $\sim\SI{9}{\giga\hertz}$, respectively. An intrinsic $f_{\rm t,i}\sim\SI{30}{\giga\hertz}$ has been calculated for the device in \cite{KocDun09}. The device has a gate length of \SI{700}{\nano\meter} and channel width of \SI{100}{\micro\meter}, its channel incorporates a large number of aligned SW-CNTs grown by CVD with a nanotube density of 5 $\rm CNTs/\mu\rm  m$ and a s:m ratio of 3:1. Metallic contacts were made from Ti/Au by a photolithography process in a ground-signal-ground (GSG) configuration. The transistor has a $p$-like behavior and the chosen bias point was $V_{\rm DS}=\SI{-1}{\volt}$ and $V_{\rm GS}=\SI{0}{\volt}$. Measurements on test ``open" structures have been reported in \cite{KocDun09}.   

\begin{figure}[!htb]
	\centering
	{\includegraphics[height=0.25\textwidth]{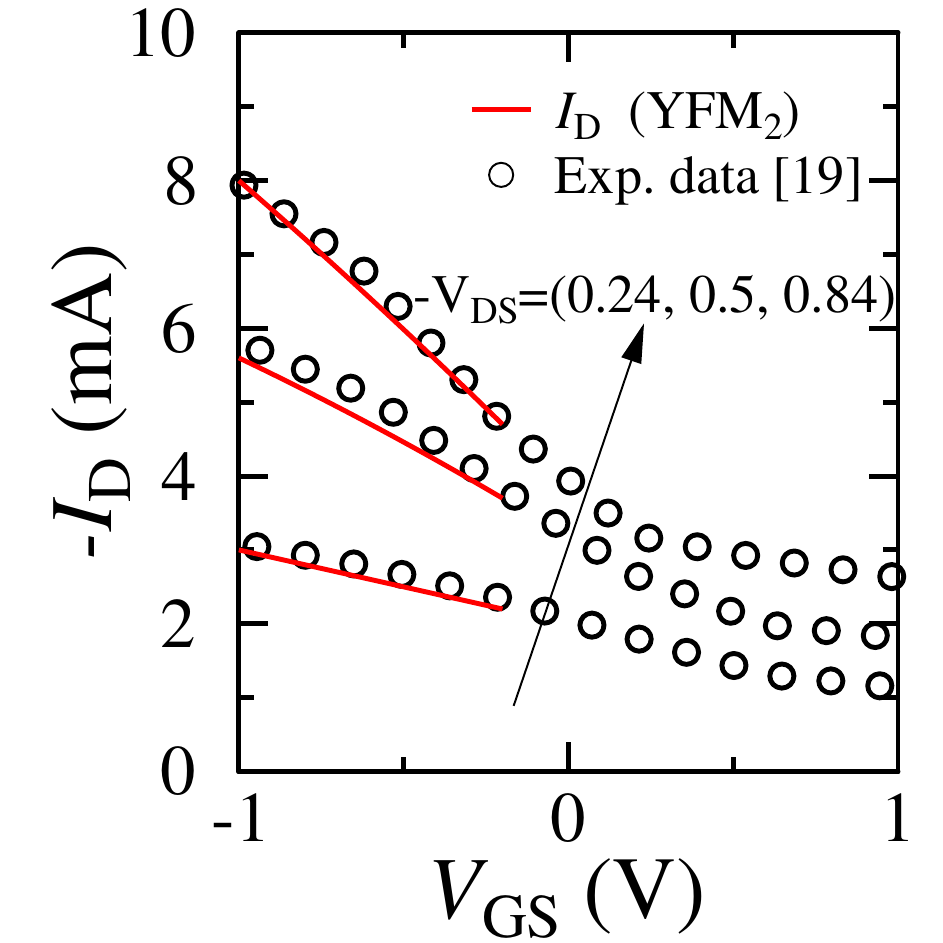}}
	\caption{Transfer characteristics extracted from output measured characteristics of a MT-CNTFET reported in \cite{KocDun09} (black markers) and $I_{\rm D}$ calculated using the extracted values from YFM$_{2}$ considering $R_{\rm C}\sim\SI{74.9}{\ohm}$ (solid red lines).}
	\label{Rc_Koc}
\end{figure}

Contact resistances have been extracted from transfer characteristics obtained using reported measured output characteristics \cite{KocDun09}. It is important to notice that the value of contact resistance has not been reported and hysteresis has been neglected in \cite{KocDun09}. Fig. \ref{Rc_Koc} shows transfer characteristics of the device and the $I_{\rm D}$ calculated with the YFM$_2$ method \cite{PacCla16}. The extracted value of contact resistance was $R_{\rm C}\sim\SI{74.9}{\ohm}$ and the contribution of each contact has been considered as identical, i.e., $R_{\rm dc}=R_{\rm sc}=R_{\rm C}/2$.

\begin{figure}[!htb]
	\centering
	\subfloat[]{\includegraphics[height=0.24\textwidth]{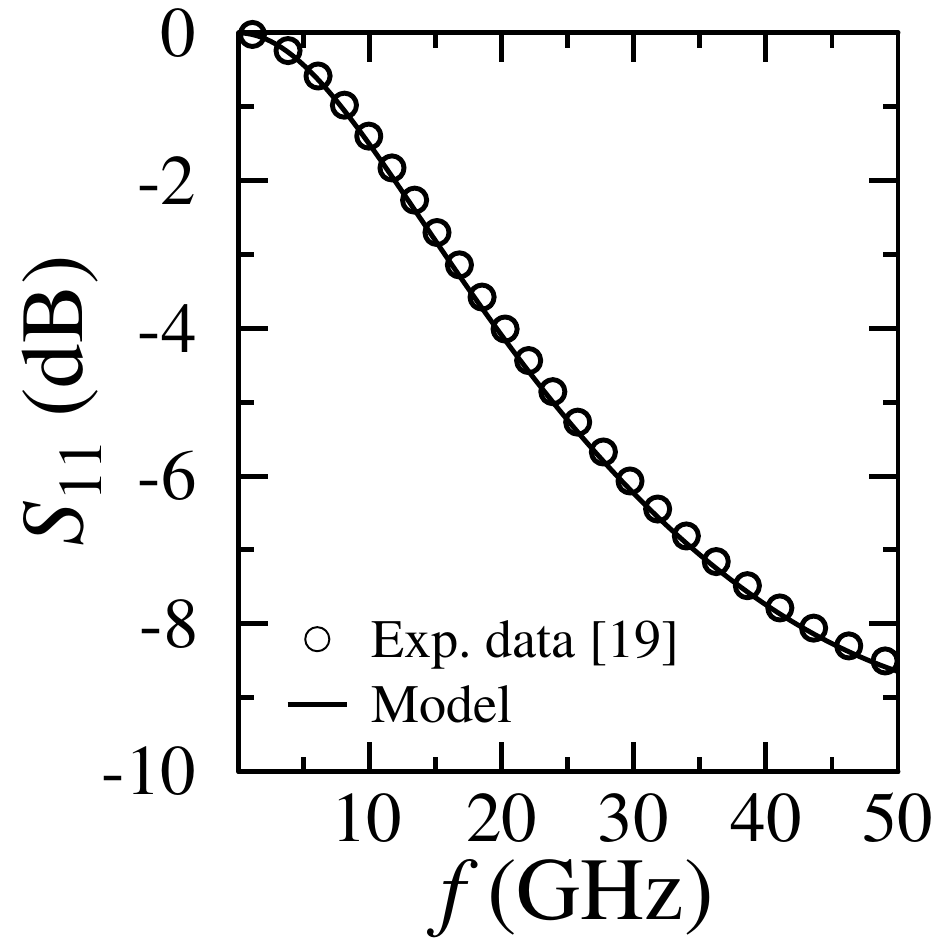}}
	\subfloat[]{\includegraphics[height=0.24\textwidth]{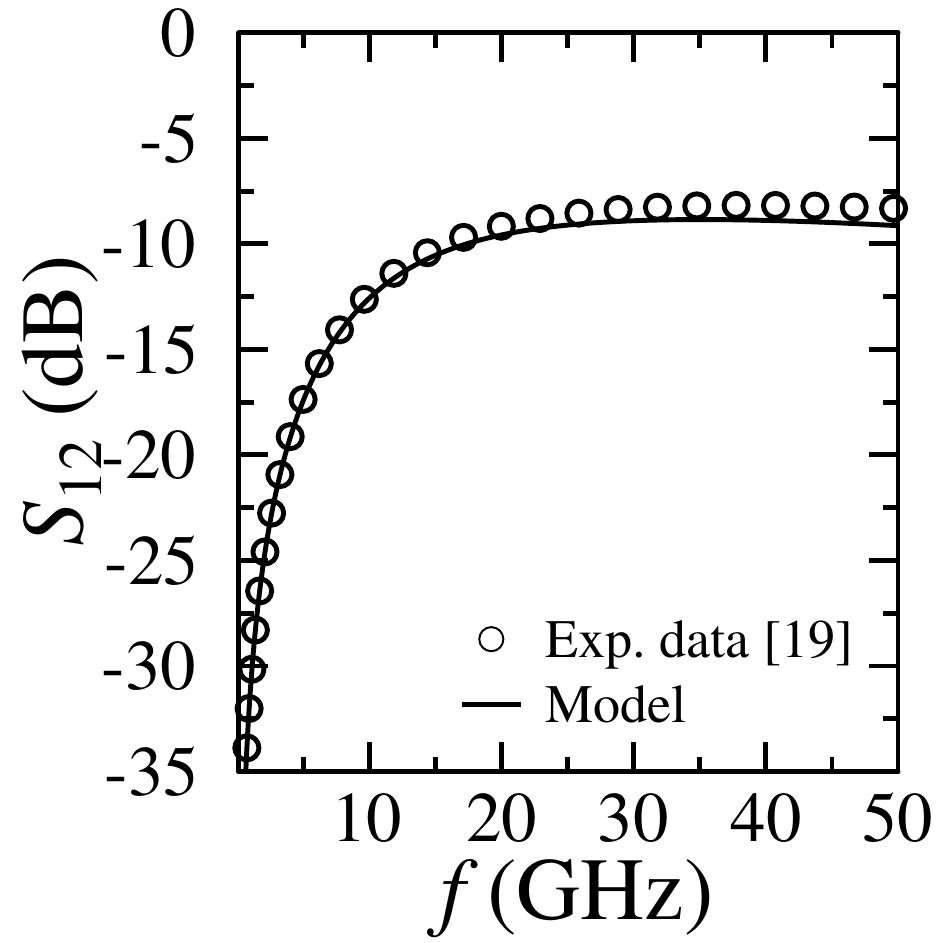}}\\
	\subfloat[]{\includegraphics[height=0.24\textwidth]{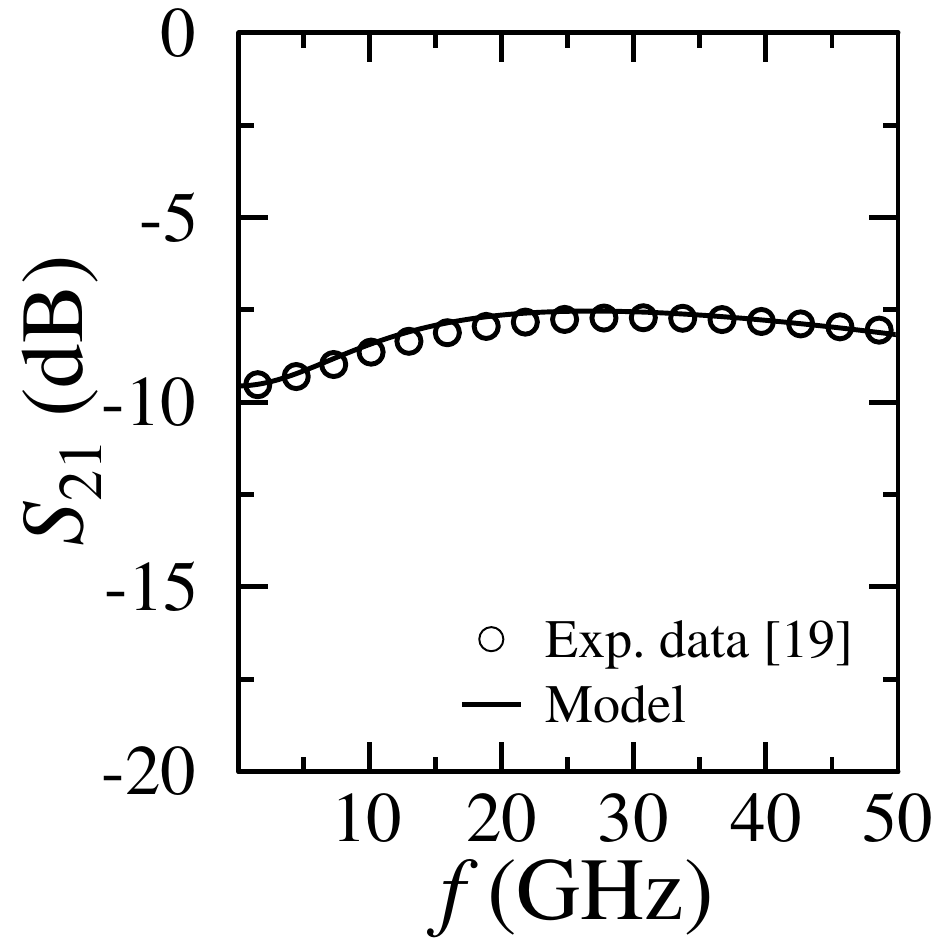}}
	\subfloat[]{\includegraphics[height=0.24\textwidth]{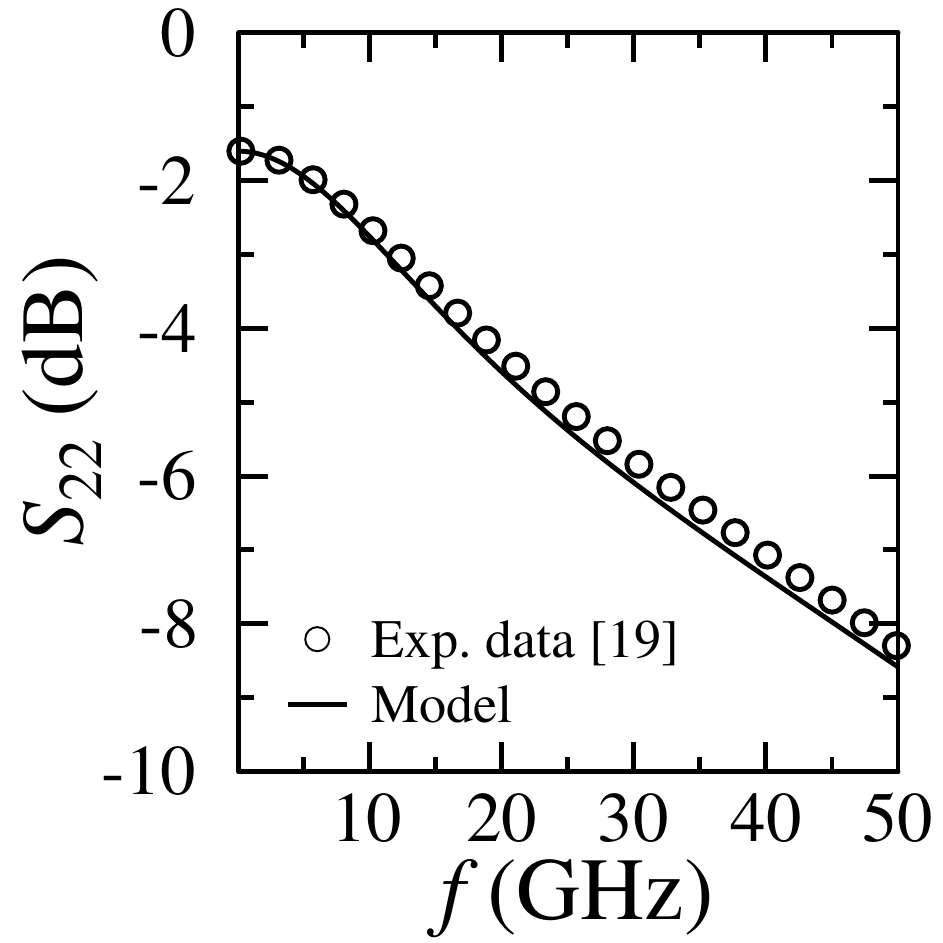}}
	\caption{Comparison of the $S$-parameters from measurements in \cite{KocDun09} and obtained with the optimized equivalent circuit. The bias point is $V_{\rm DS}=\SI{-1}{\volt}$ and $V_{\rm GS}=\SI{0}{\volt}$.}
	\label{S_Koc}
\end{figure}

A comparison between the $S$-parameters obtained from the proposed model and experimental data from \cite{KocDun09} is shown in Fig. \ref{S_Koc}. A good agreement has been achieved using the proposed methodology and the optimizing process in which only four extrinsic parameters have been fitted. In contrast with the equivalent circuit in \cite{KocDun09}, whose parameters have been extracted only by fitting the data and no extrinsic parameters have been treated, in this work an extraction procedure for both the  intrinsic and the extrinsic parameters along contact resistances, has been considered. Table ~\ref{Tab_2} shows extracted values and a comparison with the equivalent circuit values in \cite{KocDun09}.

\Table{\label{Tab_2}Comparison of the extracted and optimized parameters of the proposed equivalent circuit and reported equivalent circuit parameters for the reference fabricated device \cite{KocDun09}}
\br
&Extracted&Optimized&Reported\\
Parameter&value&value&value \cite{KocDun09}\\
\mr
\centre{4}{Intrinsic part}\\
\mr
\ns
$g_{\rm m}$	(\SI{}{\milli\siemens})	&   \SI{5.32}{}    &   \SI{5.32}{} & \SI{3.8}{}   \\ 
$g_{\rm ds}$ (\SI{}{\milli\siemens})&  \SI{2.69}{} &  \SI{2.69}{}  & \SI{1.92}{}  \\ 
$C_{\rm gd}$ (\SI{}{\femto\farad})&  \SI{11.90}{}     &  \SI{11.90}{} &  \SI{52}{}  \\ 
$C_{\rm gs}$ (\SI{}{\femto\farad})&  \SI{11.90}{}     & \SI{11.90}{}  &   \SI{65}{}    \\
$C_{\rm ds}$ (\SI{}{\femto\farad})&  \SI{17.75}{}     &  \SI{17.75}{} &  \SI{15}{}  \\
\mr
\centre{4}{Contact resistance}\\
\mr
$R_{\rm sc}$ (\SI{}{\ohm})	& \SI{37.45}{}      &  \SI{37.45}{} & -   \\
$R_{\rm dc}$ (\SI{}{\ohm})	& \SI{37.45}{}      &  \SI{37.45}{}  & - \\ 
\mr
\centre{4}{Extrinsic part}\\
\mr
$R_{\rm ga}$ (\SI{}{\ohm})	& \SI{16}{}  &  \SI{16}{}& \SI{16}{}      \\ 
$R_{\rm sa}$ (\SI{}{\ohm})	& \SI{0}{}   &  \SI{0}{}&  \SI{0}{}   \\ 
$R_{\rm da}$ (\SI{}{\ohm})	& \SI{24}{}  &  \SI{24}{}&    \SI{24}{}   \\ 
$C_{\rm gdp1}$ (\SI{}{\femto\farad})& -  &  -& -     \\ 
$C_{\rm gdp2}$ (\SI{}{\femto\farad})& \SI{68}{}      &  \SI{40}{}&  -   \\ 
$C_{\rm dsp1}$ (\SI{}{\femto\farad})& -     &  -&  -   \\ 
$C_{\rm dsp2}$ (\SI{}{\femto\farad})& \SI{12}{}      &   \SI{10}{} & -   \\ 
$C_{\rm gsp1}$ (\SI{}{\femto\farad})& -    &  -& -   \\ 
$C_{\rm gsp2}$ (\SI{}{\femto\farad})& \SI{75}{}      &  \SI{55}{} &  -  \\
$L_{\rm ga}$	(\SI{}{\pico\henry})& \SI{50}{}      &  \SI{50}{} & \SI{50}{} \\
$L_{\rm da}$	(\SI{}{\pico\henry})& \SI{50}{}      &  \SI{37}{} &  \SI{50}{}  \\
$L_{\rm sa}$	(\SI{}{\pico\henry})& \SI{0}{}      &  \SI{0}{} &  \SI{0}{}     \\
\br
\end{tabular}
\end{indented}
\end{table}

\begin{figure}[!htb]
	\centering
	\subfloat[]{\includegraphics[height=0.24\textwidth]{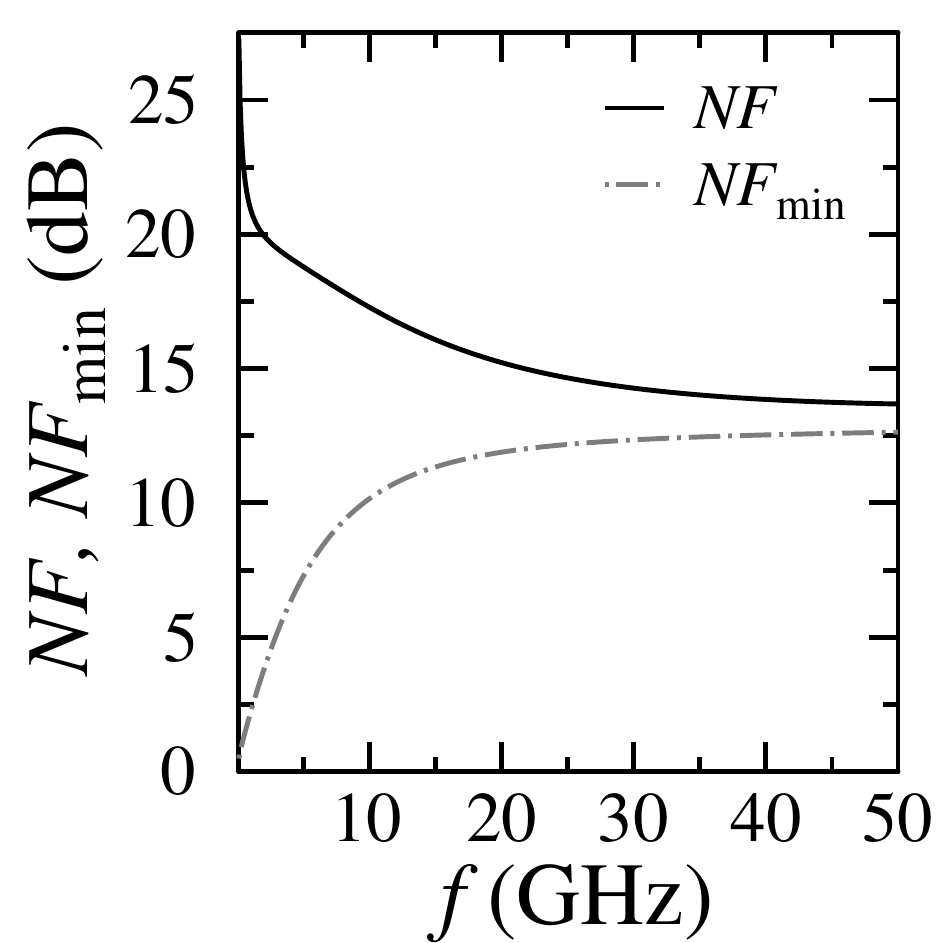}}
	\subfloat[]{\includegraphics[height=0.24\textwidth]{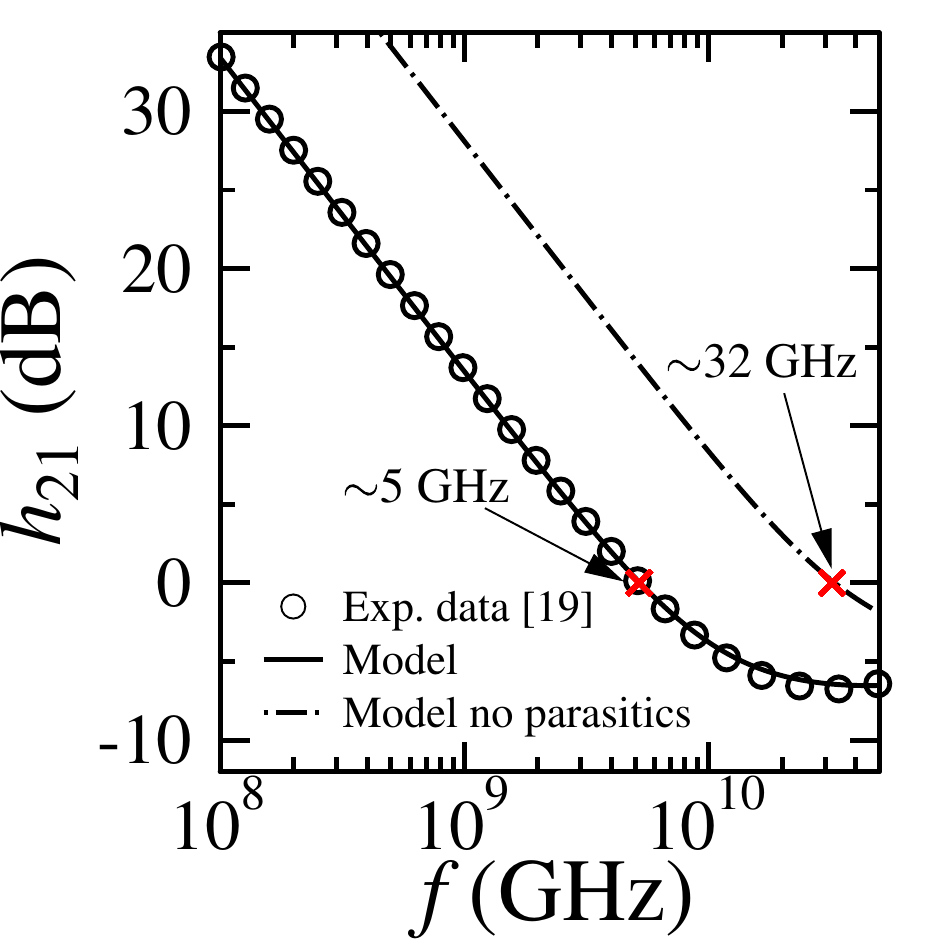}}
	\caption{(a) $NF$ and $NF_{\min}$ of the device reported in \cite{KocDun09} obtained with the proposed noise model and (b) comparison of the $h_{21}$ between experimental data from \cite{KocDun09} and the equivalent circuit and equivalent circuit without extrinsic parameters. The bias point is $V_{\rm DS}=\SI{-1}{\volt}$ and $V_{\rm GS}=\SI{0}{\volt}$.}
	\label{NF_NF_min_h21}
\end{figure}

Moreover, the noise model provides the projection of the device noise performance as shown in Fig. \ref{NF_NF_min_h21}(a). Experimental noise data reported in \cite{SakSch11} for a device from the technology described in \cite{SchKol11} indicates similar noise results. Both devices have been biased in the active transistor region. Reference \cite{SakSch11} reported a measured $NF_{\rm min}=\SI{3.5}{\decibel}$ at \SI{1}{\giga\hertz} and the proposed model based on experimental data from \cite{KocDun09} shows a $NF_{\rm min}=\SI{2.2}{\decibel}$ at the same frequency. The noise performance improvement of \cite{KocDun09} in comparison to experimental data from \cite{SakSch11} can be related to different fabrication processes, e.g., the device reported in \cite{SakSch11,SchKol11} has a channel formed by misaligned CNTs, whereas the device in \cite{KocDun09} has a channel formed by aligned CNTs.

In addition, Fig. \ref{NF_NF_min_h21}(b) shows $h_{21}$ for both: experimental data from \cite{KocDun09} and equivalent circuit. In both cases the cut-off frequency is close to \SI{5}{\giga\hertz}. Studies considering only the intrinsic part and the contact resistances of the device, i.e., without extrinsic parameters, have been made and the corresponding results can be observed in Fig. \ref{NF_NF_min_h21}(b). As it is mentioned in \cite{KocDun09} the intrinsic cut-off frequency is close to \SI{30}{\giga\hertz}. In contrast to the equivalent circuit proposed in \cite{KocDun09} in this work the intrinsic part of the equivalent circuit can reproduce the projected free-parasitic performance. 

\section{Conclusions}
\label{Sec_Con}

An equivalent circuit to model the low- and high-frequency dynamic and noise performance of CNTFETs has been proposed. The equivalent circuit allows the immediate and efficient evaluation of the performance for CNTFETs considering the effect of the contact resistance as well as extrinsic and intrinsic parameters. The proposed strategy includes the accurate extraction of the total contact resistance and a methodology to remove their contribution to the intrinsic part of the equivalent circuit. Moreover, we have shown that, noise data can be useful to perform an accurately distribution of the total contact resistance. Furthermore, a comparison of the extracted $g_{\rm m}$ and $g_{\rm ds}$ values from the proposed strategy was carried out with values from an extraction methodology for large contact resistance classic FET \cite{ChoAnt87} and similar results have been found.  

The proposed approach has been applied to two different technologies. In the former case, after the optimization process only two parasitic parameters have changed. Considering the optimized values a close behavior has been found between results from the reference compact model (CCAM) and results from the proposed equivalent circuit for $S$-parameters and noise figures of merit. In the second case, the methodology has been applied to experimental data from a top-gated CNTFET \cite{KocDun09}, in this case a good agreement between model and experimental data has been found for $S$-parameters after the optimization process in which only four extrinsic parameters have changed. Moreover, noise performance has been simulated giving insights about the expected performance of the device. The intrinsic part of the proposed equivalent circuit for this technology is also able to reproduce closely the expected performance of the device if parasitics were not considered.    

Finally, the proposed methodology can be applied to other emergent technologies with large contact resistance because this approach properly considers the effect of the contact resistance in device performance. Moreover a procedure to obtain and remove its contribution in order to extract accurately the intrinsic part of the equivalent circuit has been provided.  

\ack
This project has received funding from Instituto Polit\'ecnico Nacional under the  contract no. SIP/20196047 and from the European Union's Horizon 2020 research and innovation programme under grant agreements no. GrapheneCore2 785219 and from Spain's Ministerio de Ciencia, Innovaci\'on y Universidades under grant no. RTI2018-097876-B-C21 (MCIU/AEI/FEDER, UE) and from the Comunitat Emergent de graf\`e a Catalunya under grant no. 001-P-001702\_GraphCAT (RIS3CAT).

\end{document}